# Analysis of Financial News with NewsStream

Petra Kralj Novak, Miha Grčar, Borut Sluban and Igor Mozetič
Jožef Stefan Institute
Ljubljana, Slovenia
{petra.kralj.novak, miha.grcar, borut.sluban, igor.mozetic}@ijs.si

**Abstract**

Unstructured data, such as news and blogs, can provide valuable insights into the financial world. We present the NewsStream portal, an intuitive and easy-to-use tool for news analytics, which supports interactive querying and visualizations of the documents at different levels of detail. It relies on a scalable architecture for real-time processing of a continuous stream of textual data, which incorporates data acquisition, cleaning, natural-language preprocessing and semantic annotation components. It has been running for over two years and collected over 18 million news articles and blog posts. The NewsStream portal can be used to answer the questions when, how often, in what context, and with what sentiment was a financial entity or term mentioned in a continuous stream of news and blogs, and therefore providing a complement to news aggregators. We illustrate some features of our system in four use cases: relations between the rating agencies and the PIIGS countries, reflection of financial news on credit default swap (CDS) prices, the emergence of the Bitcoin digital currency, and visualizing how the world is connected through news.

**Keywords:** stream processing architecture, unstructured data, financial news, sentiment analysis, ontology-based entity recognition, visual analytics

## 1 Introduction

Tens of thousands of news articles are being published daily on the world wide web by thousands of news sources. People usually track only a small number of stories from a selected set of news sources, since the articles from different sources are redundant and the number of news is unmanageable. News aggregators group similar stories together and sort them according to their importance, making it easy to get an overview of the important stories of the day. By using the available on-line tools, it is, however, difficult to focus on specific entities and topics, search through their history, evaluate their relative importance, and observe the evolution of the context in which they appear over time.

We present NewsStream (`http://newsstream.ijs.si`), a web portal for interactive and visual exploration of news and blogs. It allows to focus on specific entities and topics from the financial domain, search through their history, estimate the overall sentiment attached to them through time, evaluate the relative importance of events, summarize the news through tag clouds, and observe the evolution of the context in which they appear. The main characteristic of this portal is in its focus on when something happened, how big was the story and its sentiment ahead of what the event was. By doing so, the NewsStream portal gives a novel perspective on the reporting in global news media. Another unique aspect of NewsStream is its focus on the financial domain with lightweight semantics and sentiment analysis.

In three use cases, we exemplify different features of the querying and visualization mechanisms and hint at a variety of investigations than could be further performed by using the NewsStream portal. In the first use case, we seek for insights in the financial domain: we investigate the time-lines of news about credit rating agencies related to the European countries with increasing sovereign debt. The second use case studies the relation of economic news and their sentiment to the CDS prices of a troubled eurozone country, Cyprus. The third use case describes the emergence of Bitcoin, a controversial digital currency, which received widespread attention in 2013.

The basis of the NewsStream portal is DacqPipe, a data acquisition and processing pipeline. DacqPipe consists of a number of connected data analysis components, each performing some text processing operation. Depending on the load, several pipelines can be run in parallel. Most of the DacqPipe components perform language-independent text processing with the exception of the semantic annotation component which is tailored to the financial domain by employing a domain-specific ontology of financial entities and terms. The result of DacqPipe are two databases: a set of annotated documents stored on the file system, and a dedicated database which supports efficient querying over specific, finance-oriented aspects of the news. The databases can be queried through an API, which provides views, aggregates, and access to the documents. The applicability of the data collected and processed by DacqPipe is very diverse [20, 21, 5, 1]; one of the applications is the NewsStream web portal.

This report is organized as follows. Section 2 is dedicated to the description of the architecture and details of the DacqPipe data acquisition and processing pipeline. Section 3 presents the main functionalities of the NewsStream portal which are then illustrated in three use cases. In Section 4, we provide an overview of the related work, the relationship between NewsStream and other news aggregation web sites, emphasize similarities and differences to those approaches and present the complementarity and added value of NewsStream. Section 5 summarizes the presented work and gives directions for further development and research.

## 2 Data Acquisition and Processing

Our DacqPipe data acquisition and processing pipeline consists of a number of interoperating technologies. It is responsible for acquiring unstructured data from several data sources, preparing it for the analysis, and brokering it to the appropriate analytical components. DacqPipe is running continuously since October 2011, polling the Web and proprietary APIs for recent content, turning it into a stream of preprocessed text documents. It is composed of two main parts: (i) data acquisition and (ii) semantic data processing. The pipeline is schematically presented in Figure 3.

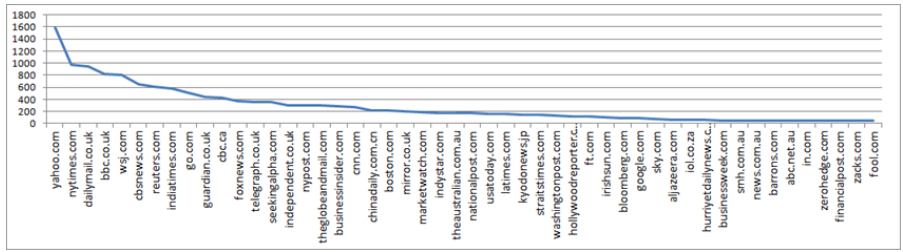

Figure 1: The fifty most productive domains with their daily average document production.

## 2.1 The Data Acquisition Pipeline

The news articles and blogs are collected from 2,600 RSS feeds from 175 English language web sites, covering the majority of web news in English and focusing on financial news and blog sources. We collect data from the major news providers and aggregators (such as yahoo.com, dailymail.co.uk, nytimes.com, bbc.co.uk, wsj.com) and also from the most relevant financial blogs (such as zerohedge.com). The fifty most productive web sites account for 80% of the collected documents (see Figure 1).

Our data acquisition started on October 24, 2011. In the period from October 2011 to January 2014, more than 18 million unique documents were collected and processed. The number of acquired documents per day is presented in Figure 2. On an average work day, about 18,000 articles are collected. The number is substantially lower during weekends; around 10,000 per a weekend day. Holidays are also characterized by a lower number of documents.

Content from news, blogs, forums, and other web content is not immediately ready to be processed by the text analysis methods. Web pages contain a lot of 'noise' or 'boilerplate' (i.e., undesired content such as advertisements, copyright notices, navigation elements, and recommendations) that needs to be identified and removed before the content can be analyzed. For this reason, DacqPipe contains various data cleaning components. In general, DacqPipe consists of (i) data acquisition components, (ii) data cleaning components, (iii) natural-language preprocessing components and (iv) semantic annotation components.

The data acquisition components are RSS readers that poll for data in parallel. One RSS reader is instantiated for each web site of interest. The RSS sources, corresponding to a particular web site, are polled one after another by the same RSS reader to prevent the servers from rejecting requests due to concurrency. An RSS reader, after it has collected a new set of documents from an RSS source, dispatches the data to one of several processing pipelines. The pipeline is chosen according to its current load which ensures proper load balancing. A processing pipeline consists of a boilerplate remover, duplicate detector, language detector, sentence splitter, tokenizer, part-of-speech tagger, lemmatizer, stop-word detector and a semantic annotator. Some of the components are based on LATINO, our own text mining library [10], while others use the functionality from the OpenNLP library [2]. In the following, we briefly present the data processing components employed in DacqPipe.

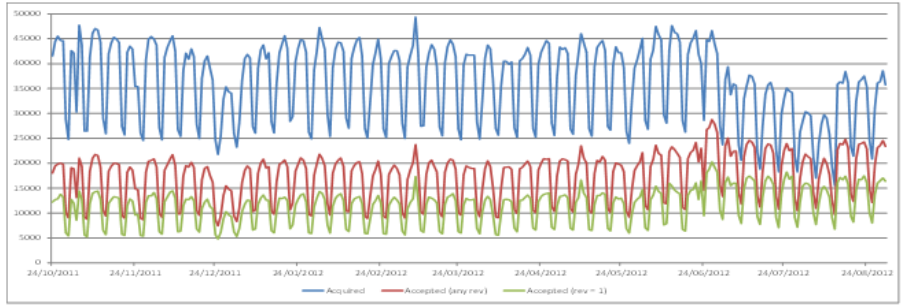

Figure 2: The number of acquired documents per day. The blue color represents all acquierd documents, the red line represents unique documents, the green line represents first revisions of documents.

*Boilerplate Remover.* Extracting meaningful content from Web pages presents a challenging problem which was already extensively addressed in the offline setting. Our setting, however, requires content extraction from streams of HTML documents in real-time. The content extraction algorithm is efficient, unsupervised, and language-independent. It is based on the observation that HTML documents from the same source normally share a common template. The core of the proposed content extraction algorithm is a data structure called the URL Tree. The performance of the algorithm was evaluated in a stream setting on a time-stamped semi-automatically annotated dataset which was also made publicly available [26]. It was shown that the algorithm clearly outperforms an extensive set of open-source boilerplate removal algorithms.

*Duplicate Detector.* Due to news aggregators and redirect URLs, one article can appear on the web with many different URLs pointing to it. To have a concise dataset of unique articles, we developed a duplicate detector that is able to determine whether the document was already acquired or not.

*Language Detector.* By using a machine learning model, the language detector recognizes the language and discards all the documents that are detected to be non-English. The model is trained on a large multilingual set of documents. The basic features for the model training are frequencies of several consecutive letters.

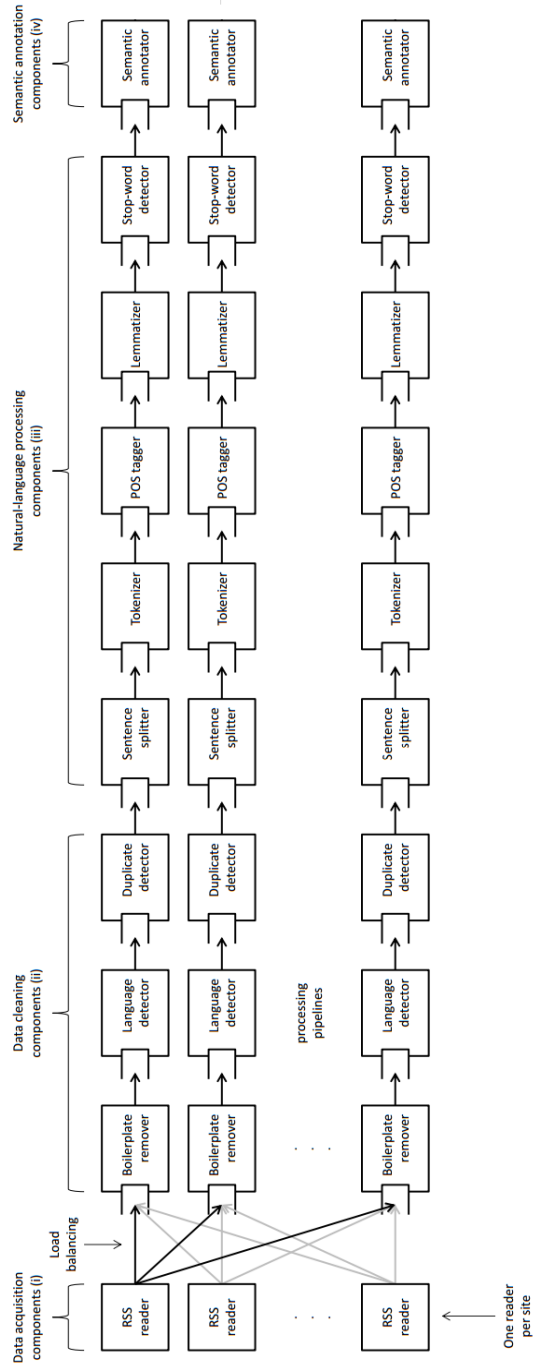

Figure 3: The data acquisition, processing and semantic annotation pipeline.

*Sentence Splitter.* The sentence splitter splits the text into sentences. The result is the input to the part-of-speech tagger. We use the OpenNLP [2] implementation of the sentence splitter.

*Tokenizer.* Tokenization is the process of breaking a text into words, phrases, symbols, or other meaningful elements called tokens. In DacqPipe, we use our own implementation of tokenization, which supports the Unicode character set and is based on rules.

*Part-of-Speech Tagger.* The part-of-speech (POS) tagger marks tokens with their corresponding word type (e.g., noun, verb, proposition) based on the token itself and the context of the token. The part-of-speech tagger from the OpenNLP library [2] is used.

*Lemmatizer.* Lemmatization is the process of finding the normalized forms of words appearing in text. It is a useful preprocessing step for a number of language engineering and text mining tasks, and especially important for languages with rich inflectional morphology. In our data acquisition pipeline, we use LemmaGen [13], the most efficient publicly available lemmatizer trained on large lexicons for various languages.

*Stop-word Detector.* In automated text processing, stop words are words that do not carry any semantic meaning. DacqPipe detects and annotates such words.

## 2.2 Semantic Data Processing

DacqPipe's data acquisition, cleaning and natural-language preprocessing components are domain-independent and biased towards finance only by the selection of RSS sources. Semantic data processing is, however, tailored to the financial domain by employing a domain-specific lightweight ontology of financial entities and terms. The ontology also includes gazetteers, and a dictionary of positive and negative words for dictionary-based sentiment analysis. A gazetteer specifies the lexicographic information about possible appearances of an entity in text, and is used for named entity recognition (NER).

### 2.2.1 Ontology of Financial Entities and Terms

The ontology consists of three main categories: (i) financial entities, (ii) financial terms related to the latest financial crisis, and (iii) geographical entities. Most of the ontology was automatically constructed by integrating various data sources. The geographical entities (continents, countries, cities, organizations (such as European Union and United Nations) and currencies) and the relations between them were extracted from GeoNames (`http://www.geonames.org/`). We used

MSN Money (http://money.msn.com/) to link stocks to the companies that issue them. We added a list of 'over-the-counter' stocks from OTC Markets (http://www.otcmarkets.com/home). The hierarchy of financial terms related to the financial crisis was developed in collaboration with economy experts. It includes the main European politicians and economy leaders (called "protagonists"), central banks and other financial institutions, rating agencies, and fiscal and monetary policy terms. The class hierarchy of the ontology and the number of concepts per class are given in Figure 4.

### 2.2.2 Semantic Annotator

Each entity in the ontology is associated with a gazetteer; a gazetteer specifies rules which define possible appearances of the entity in text. For example, 'The United States of America' can appear in text as 'USA', 'US', 'The United States' and so on. The rules include capitalization, lemmatization, and POS tag constraints, 'must contain' constraints (i.e., term from another gazetteer must be detected in the document or in the sentence) and 'followed by' constraints. An example of a gazetteer attached to an entity is given in Figure 5.

### 2.2.3 Sentiment Analysis

Sentiment analysis is the computational study of how opinions, attitudes, emotions, and perspectives are expressed in language. There exist three main approaches to sentiment analysis: rule-based, classification-based and the dictionary-based approach. In [19], the authors provide a survey of sentiment analysis challenges and the approaches used to tackle them. Specifics of opinion mining in news articles are described in [25]. The text classification approach involves building classifiers from labeled instances of texts or sentences. It is essentially a supervised classification task. The dictionary-based approach involves calculating orientation for a document from the semantic orientation of words or phrases in the document. We use the dictionary-based approach, since we are constrained by real-time performance and do not have a corpus of sentiment labeled news articles.

We use the Harvard-IV-4 Sentiment Dictionary [29, 30] with 2,006 negative words and 1,636 positive words. We compute document sentiment polarity according to Equation 1, where *pos* denotes the number of positive words in a text segment (paragraph or document) and *neg* denotes the number of negative words in this same text segment. The range of document polarity is between $-1$ and $1$, $-1$ being the most negative sentiment polarity, $1$ the most positive sentiment polarity and $0$ indicating neutral sentiment.

$$polarity = \frac{pos - neg}{pos + neg} \quad (1)$$

For efficiency reasons, we have precomputed document sentiment polarity for each of the stored documents and for each paragraph in the documents. We compute the sentiment per entity per day by aggregating the sentiment

```
Company (103)
- Fund (114)
- Industry (533)
- Bank (85)
- Insurance (37)
- Other (152)
Geographical region (32)
- City (37)
- - Capital city (243)
- Country (239)
- - PIIGS Country (5)
- - BRICS Country (5)
- Continent (7)
Organization (29)
Eurocrisis
- Protagonist (28)
- Finance
- - Financial_Instrument (17)
- - - Over the counter (1572)
- - - Stock (977)
- - - Currency (15)
- - - Digital_Currency (5)
- - - - Digital_Currency_Market (4)
- - - Index
- - - - Stock Index (18)
- - Financial_Markets
- - - Financial_markets_term (9)
- - - Neg_Financial_Sentiment (2)
- - - Financial_Crisis_Term (5)
- - - Pos_Financial_Sentiment (4)
- - Financial_Institution
- - - EU_Mechanism (11)
- - - EU_Financial_Institution (8)
- - - Rating_Agency (4)
- - - Other_Financial_Institution (3)
- - - Int_Financial_Institution
- - - US_Financial_Institution (3)
- - - Central_Bank (3)
- - Public_Finance
- - - Monetary_Policy_Term (8)
- - - Fiscal_Policy
- - - - Workforce_term (9)
- - - - Public_Spending
- - - - - Loan_Risk_Term (5)
- - - - - Budget_Term (17)
- - - - - Loan_Term (6)
- - - - - Financial_Failure_Term (6)
- - - - - Public_Debt_Term (4)
- - - - - Loan_Insurance_Term (3)
```

Figure 4: The ontology of classes of financial objects with the number of instances (entities and terms) belonging to each class.

```
:bank_Banca_Monte_Dei_Paschi_Di_Siena
    :identifiedBy :bank_Banca_Monte_Dei_Paschi_Di_Siena_Gazetteer;
    :identifiedBy :bank_Banca_Monte_Dei_Paschi_Di_Siena_ACR_Gazetteer;
    a :Bank;
    RDFS:label "Banca Monte Dei Paschi Di Siena".

:bank_Banca_Monte_Dei_Paschi_Di_Siena_Gazetteer
    :term "Banca Monte Dei Paschi Di Siena";
    :term "Monte Dei Paschi";
    a :Gazetteer.

:bank_Banca_Monte_Dei_Paschi_Di_Siena_ACR_Gazetter
    :term "MPS/c=auc";
    :term "BMPS/c=auc";
    :hasDocumentLevelCondition :gaz_bank;
    a :Gazetteer.
```

Figure 5: An example of a semantic entity and its gazetteer, encoded as triples in the Notation 3 (or N3) format.

polarity of documents where an entity appears in a certain day. The aggregation function can either be *sum* (default) or *average*. The sentiment granularity level (sentiment scope) can be either document or paragraph.

By using this approach to sentiment analysis, we are able to automatically detect sentiment regarding an entity in real-time, and aggregate it on a daily basis.

*Semantic Annotation Database.* The information about the terms (entities and sentiment words) and their location (paragraph, character position) in each document is stored in an SQL database. Additionally, the entity-class relationship and the hierarchy of ontology classes is also stored. Various meta-data about the document, including the document title, acquisition and publication time, source domain, response URL, allow the drill-down to the concrete document. Some aggregates, such as the sentiment polarity of each document and each paragraph, are also precomputed and stored in the database for performance reasons. Such a database enables efficient and diverse querying.

We have implemented an API for data access. The API allows us to retrieve document titles and response URLs, documents by sources, dates, recognized entities, sentiment words and aggregated sentiment over documents/paragraphs. We use these and several other queries in our applications.

The use of the data collected and processed by DacqPipe is very diverse. There are already several applications which rely on the DacqPipe API such as (i) a new measure of cohesiveness of news, the News Cohesiveness Index (NCI) [20], (ii) a method for constructing time varying co-occurrence networks of entities from texts [21], (iii) a decision support model for estimating reputation risk of a bank, based on the financial properties of the products it offers and the sentiment about their counterparts [5], (iv) the detection of illegal 'pump and dump' scenarios, where a low volume stock is heavily advertised (to pump its price) and then sold (dumped) with a profit [1].

The rest of this paper is dedicated to a new application: the NewsStream web portal for news and blogs analytics. The portal allows everybody to easily access and query our data.

# 3 The NewsStream Portal

NewsStream (`http://newsstream.ijs.si`) is a web-based interactive news and blogs analytics portal. Prior to *what* happened, it focuses on the questions (i) when something important happened, (ii) what was the amplitude (volume), and (iii) what was the related sentiment. Its main functionality is the projection of relevant documents (news articles and blog posts) and their sentiment onto a time-line, summarization of the documents by aggregation, and drill-down to the original documents. The main features of NewsStream are:

- Visualization of occurrences and co-occurrences of financial and geographical entities and financial terms, as lines, stacked lines or canyon flow charts;

- Sentiment aggregation (either sum or average) in different scopes (document or paragraph);

- Retrieval of documents strictly related to finance;

- Adjustable window size for smoothing;

- Visualization of tag clouds that summarize the content of the documents for a selected day and entity (or entity pair);

- Drill-down to the individual annotated documents with hyper-links to the originals ordered by relevance to the query or by sentiment;

- Computation of trends with parametrized MACD (moving average convergence/divergence);

- Adjustable confidence level with min-occurrence per document;

- Intuitive and friendly user interface with drop-downs and automated query completion.

- Visualization of temporal country co-occurrence networks extracted from news and embedded on the world map.

These atomic operations can provide insight and help answering more profound questions. The basic use case of NewsStream is to visualize the number of documents and aggregated sentiment of documents for selected entities or entity co-occurrences in time. Abnormalities in the fluctuation of the number of documents reflect an entity or co-occurrence of entities being unusually present in the news. The respective aggregated sentiment reflects the positive or negative context in which the entity (or entities) occur. The contents summary by tag clouds is used to quickly gain insight in an event while the drill-down to the specific articles allows in-depth investigations. The visualization of country co-occurrences in news on the world map enables to explore how the world is connected in the news over time. All this composes a unique tool that helps to quickly reveal the story of a selected entity or entities in time, as reported by the English-language media in history as well as in near real-time.

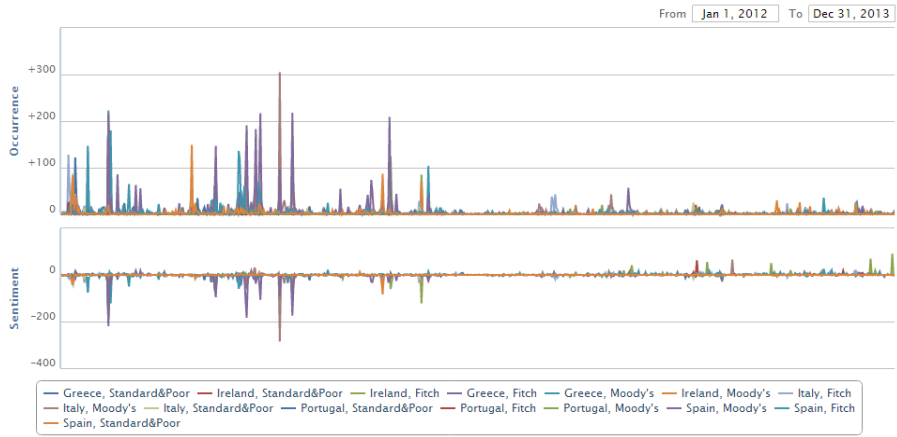

Figure 6: The PIIGS countries co-occurring with the rating agencies Fitch, Standard&Poor's and Moody's.

In the use cases, we show how visual big data analytics is made possible to domain experts (like economists and financial experts) through a view on the global news that is provided by the NewsStream portal. The four use cases exemplify different functionalities of the querying and visualization mechanisms and hint at a variety of investigations than could be further performed by using our portal.

## 3.1 Rating Agencies and the PIIGS Countries

NewsStream can be used to gain on overview of a situation. In this example, we focus on the European sovereign debt crisis in years 2012 and 2013. We investigate the news mentioning credit rating agencies and the so-called PIIGS countries—the five Eurozone countries, which were considered economically weaker at the beginning of the financial crisis. Figure 6 shows document-level co-occurrences of the PIIGS countries (Portugal, Ireland, Italy, Greece and Spain) with rating agencies (Fitch, Standard&Poor and Moody's).

At a first glance, we can see that:

- Rating agencies brought mostly bad news in year 2012, like in the case of Italy and Fitch on January 10, Spain and Moody's on June 13, and Italy and Moody's on July 13 (some of the prominent picks in the timeline);

- In the year 2013, the rating agencies co-occur with PIIGS countries infrequently compared to the year before with the biggest story being Fitch upgrading Greece credit rating on May 14, 2013 (this story was reported six times less then the downgrades one year earlier);

- More optimism is on the horizon in year 2013 since the sentiment of the reported news is predominantly positive. By using the drill-down functionality of NewsStream, we discover that the rating agency Fitch changed the outlook to stable for Greek bonds.

We compare the co-occurrences of rating agencies and selected countries with the country ratings data from CountryEconomy.com (`http://countryeconomy.com/ratings`). The credit rating changes are reported consistently. For example, the rating downgrade of Italy by Moody's on February 13, 2012 was covered by 210 articles the following day and the rating downgrade of Italy by Moody's a few months later on July 13, 2012 was reported by 305 articles on the downgrade day. Not all the rating changes have the same media attention and, surprisingly, the media attention does not seem to decrease over time. The rating downgrades are also reported more than the upgrades.

## 3.2 Reflection of Financial News in CDS Prices

In this use case, we consider the European sovereign debt crisis and ask the question: what insights can we get from the NewsStream portal about the evolution of the crisis and what are the relations to the financial markets? We show that the portal can guide one (1) to detect emerging topics through canyon-flow visualization, (2) to get a condensed overview of relevant news through tag clouds, (3) to track the evolution of the sentiment through time, and (4) to drill all the way down to individual news.

First, we focus on the financial news about the PIIGS countries (Portugal, Ireland, Italy, Greece, and Spain) and Cyprus, nicknamed PIIGSC. The relative distribution of news between the PIIGSC individual countries during the last year is in Figure 7 (note that the window size is one week, to allow for smooth transitions of occurrences). One can observe that the interest in the most problematic country, Greece, has mostly disappeared after the onset of the crisis in 2008. Also, the attention to Spain has decreased, while it has increased for Ireland. There is very few news about Portugal, with a notable jump in July 2013. Cyprus is mostly ignored until March 2013, when one observes a huge increase in the number of news concerning Cyprus.

Next, we focus on the news concerning Cyprus in conjunction with the entity 'eurozone' during March 2013. The top chart in Figure 8 shows the volume of joint occurrences of both terms in the number of paragraphs (not documents), while the bottom chart shows the sentiment associated with both terms in the same paragraph. One observes a large increase in the number of news from March 12 to March 18, an initial positive sentiment (until March 15), and then a sharp drop in sentiment until March 18.

A summary of news can be represented by a tag cloud. Figure 9 gives two tag clouds for March 15: the first one is constructed from the titles of news, while the second one is constructed from the ontology entities recognized in the bodies of the news. The size of a tag is proportional to the number of occurrences of the term in the news. Cyprus is prevailing in both clouds (no surprise, since it is part of the query). However, it is interesting to observe the difference: while titles emphasize 'bailout', the news themselves emphasize EU related terms (Brussels, Europe, European Union, eurozone), International Monetary Fund and United States.

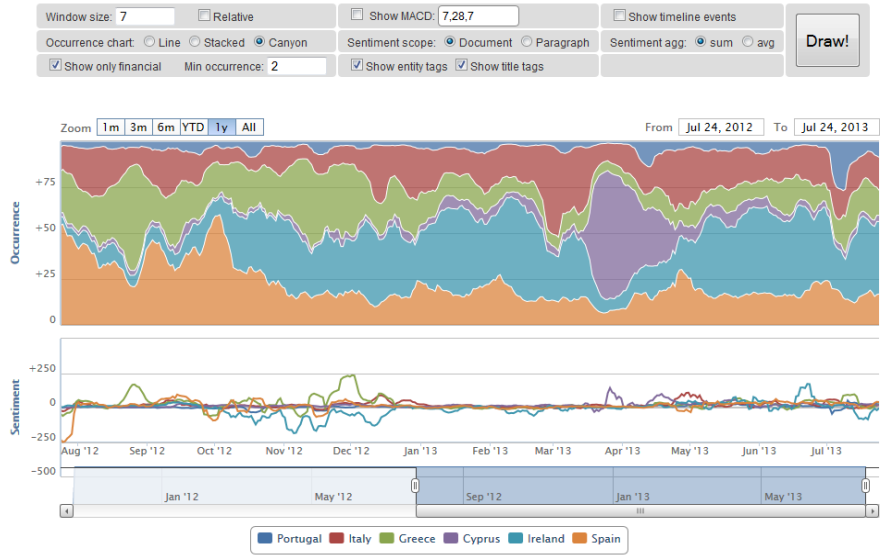

Figure 7: Canyon flow of financial news about the PIIGS countries (Portugal, Ireland, Italy, Greece, and Spain) and Cyprus.

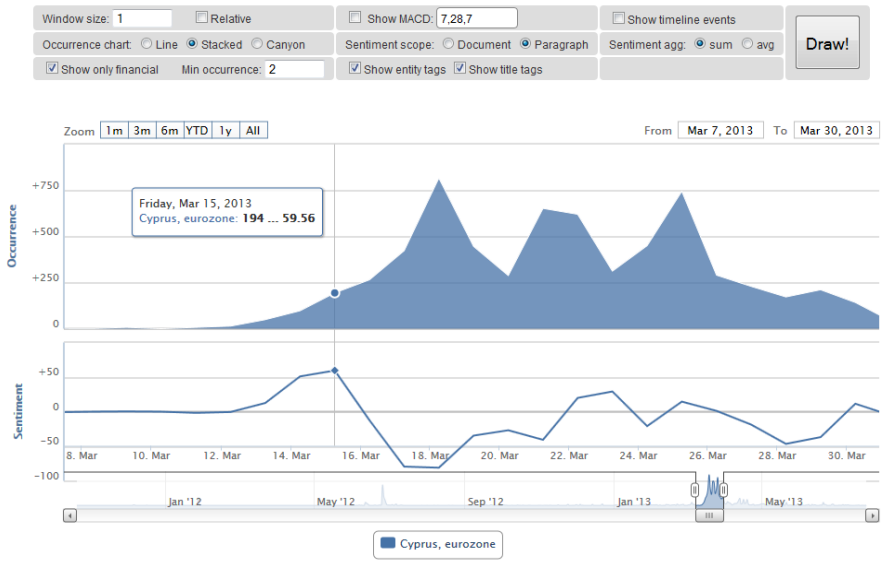

Figure 8: Top - Joint occurrences of Cyprus and 'eurozone' in the number of paragraphs. Bottom - The sentiment associated with both terms in the same paragraph.

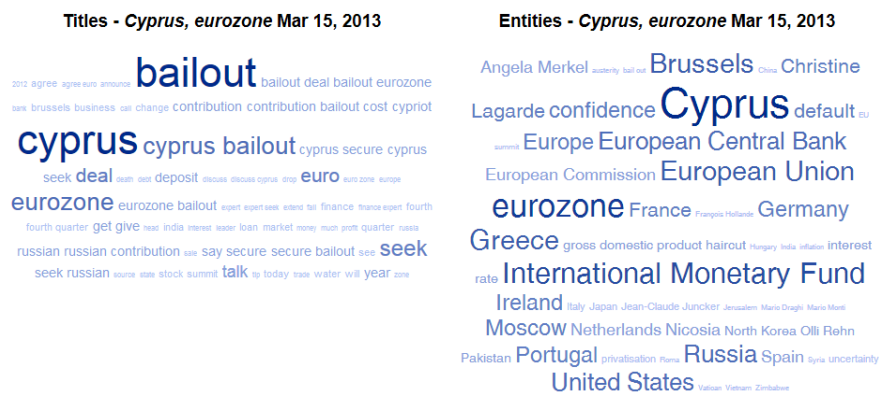

Figure 9: Left - A tag cloud of titles of documents mentioning Cyprus and eurozone. Right - A tag cloud of ontology entities recognized in the news bodies, together with Cyprus and eurozone.

The tag clouds for the subsequent days are very similar, so they offer no insight into why the sentiment about Cyprus and eurozone changed. One has to dig into the individual news for the following four days. Figure 10 gives the three most positive news for March 15, the three neutral news for March 16, and the three most negative news for March 17 and 18. From the titles alone, it is obvious that there was an initial agreement on the Cyprus bailout which evoked positive sentiment in the news. However, during the weekend (March 16 and 17) more details of the deal were revealed which culminated in the sentiment drop on March 18.

It is interesting to observe what was going on in the financial markets at the same time. The perception of financial markets about an individual country is usually reflected in the price of its sovereign bonds or CDS (credit default swap). Essentially, a CDS is insurance for the buyer of the country bonds in the case that the country is unable to repay its debt (declares a default). The higher price of CDS reflects the perception of the higher risk of default. We compare the sentiment about Cyprus to the closing price of CDS in the middle of March 2013 (see Table 1). CDS prices are given in basic points (b.p.), i.e., 100 b.p. equals 1% of the underlying bond value.

Until the emergence of Cyprus in the media (March 15), the CDS prices were stable (albeit relatively high in comparison to the other eurozone countries). After the weekend (March 16 and 17) of negative news, the closing price on Monday (March 18) jumped for 21%. A similar jump of 15% occurred on the next day (March 19), and then the price stabilized again. From this isolated observation, one cannot conclude that bad news alone influence financial markets. However, a deeper insight into the news, and especially the related sentiment, gives a hint at what might happen in the future.

| | | | |
|---|---|---|---|
| 1. | Eurogroup head: political agreement on 10 bln euro Cyprus bailout **0.520** | 43. | German finmin asks Bundestag to agree Cyprus deal **0.024** |
| 2. | Entering the final stretch to bailout deal **0.425** | 44. | The Meaning of Cyprus **0.022** |
| 3. | Eurozone discussing Cyprus bailout **0.400** | 45. | A look at Cyprus' decision to tax depositors **0.020** |
| 163. | Euro Watch Asian Markets Drop on Latest Euro Concerns **-0.447** | 371. | Market Falls on Fear Cyprus Deal Could Hurt Euro **-0.489** |
| 164. | Run on banks in Cyprus adds to eurozone woes **-0.452** | 372. | Cyprus bank levy spooks Hong Kong investors **-0.512** |
| 165. | Run on banks in Cyprus adds to eurozone woes **-0.491** | 373. | Stock indexes lower on Cyprus fears **-0.519** |

Figure 10: The three most positive news for March 15 (top left), three neutral news for March 16 (top right), and the three most negative news for March 17 (bottom left) and 18 (bottom right) about Cyprus and eurozone.

Table 1: The CDS closing price for Cyprus (in basic points).

| Date | CDS price | daily change |
|---|---|---|
| 3/14/2013 | 573.025 | - |
| 3/15/2013 | 573.468 | 0.1% |
| 3/18/2013 | 697.108 | +21.6% |
| 3/19/2013 | 802.214 | +15.1% |
| 3/20/2013 | 819.877 | +2.2% |
| 3/21/2013 | 819.877 | 0% |

### 3.3 Emergence of Bitcoin

Bitcoin is the world's first decentralized digital currency. It was proposed in 2009 [18], but has been increasingly gaining public attention over the year 2013, both in policy discussions [9] and in the popular press. In this use-case, we explore the news and blogs coverage of Bitcoin and its reflection to Bitcoin price, trading volume and market capitalization.

Each year, the Australian National Dictionary Centre selects a 'word of the year'. The words chosen for the shortlist are selected on the basis of having come to some prominence in the Australian social and cultural landscape during the year. The 2013 word of the year is Bitcoin. Our analysis of the documents collected in years 2012 and 2013 confirm Bitcoin and related terms to be gaining the most attention in year 2013. From more than 4000 named entities and financial vocabulary terms from our ontology, Bitcoin and digital currency related vocabulary (Mt.Gox, an exchange which allows users to trade Bitcoins for US Dollars, and other synonyms of digital currency, like cryptocurrency) got three out of top five placements (Table 2). Less than a thousand documents mentioned Bitcoin in year 2012 and more then thirty-eight thousand documents mentioned Bitcoin in 2013.

Table 2: The top five named entities and financial vocabulary terms gaining the most prominence in year 2013 compared to year 2012. #2012 and #2013 are the number of documents mentioning the entity in the respective year and last column is the ratio of the two values.

| Entity | #2012 | #2013 | Ratio |
|---|---|---|---|
| Mt.Gox | 8 | 1154 | 144.2 |
| Bitcoin | 999 | 38963 | 39.0 |
| Volgograd | 374 | 7129 | 19.1 |
| Digital currency term | 716 | 9216 | 12.9 |
| Bandar Seri Begawan | 240 | 2897 | 12.1 |

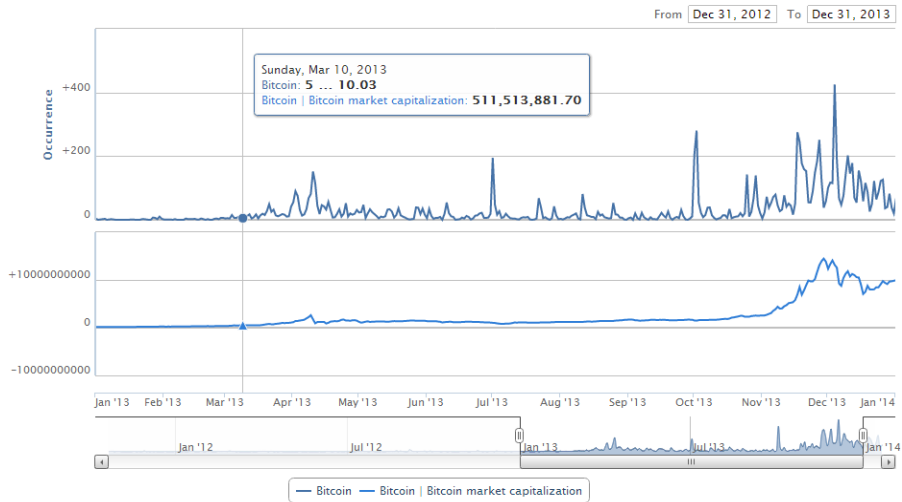

Figure 11: Bitcoin occurrence timeline and Bitcoin market capitalization in year 2013. The first significant wave of news reports about Bitcoin started just after Bitcoin has reached market capitalization of $500 Million on March 14, 2013.

The Bitcoin occurrence timeline (Figure 11, top) shows us when articles about Bitcoin stopped being solitaires but gained popularity in mainstream media. The first significant wave of news reports about Bitcoin started just after Bitcoin has reached market capitalization of $500 Million on March 14, 2013 (see Figure 11, bottom), probably contributing to the increase on the demand for Bitcoin. As Bitcoin reached the price of $200 in April and market capitalization just over $1.7 Billion on April 11, 2013, payment processors BitInstant and Mt.Gox experienced processing delays due to insufficient capacity resulting in

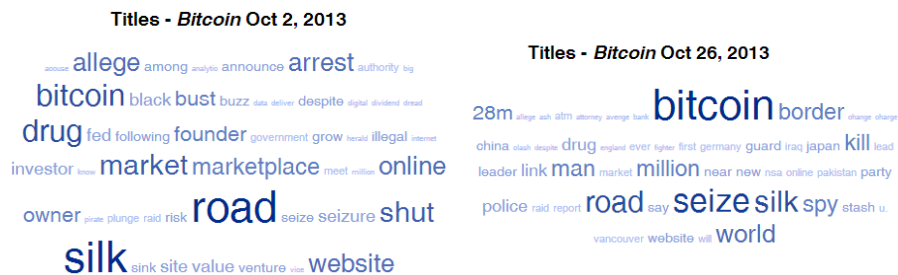

Figure 12: The three most positive (left) and three most negative (right) news on Bitcoin on April 11, 2013. The first number next to the title is the document sentiment and the second number is the occurrence count of the term 'Bitcoin' in the document.

Figure 13: Tag cloud of document titles of documents about Bitcoin on October 2 and October 26 2013.

the Bitcoin exchange rate dropping from $266 to $76 before returning to $160 within six hours. We use the NewStresm drill-down functionality to look at the documents about Bitcoin on that day. By sorting the documents by sentiment, one can easily access the most positive and negative documents of the day. Figure 12 shows the titles of the three most positive documents (top) and the titles of the most negative documents (bottom).

The Bitcoin occurrence timeline (Figure 11, top) shows when news about Bitcoin were reported. It is just a click away to see the documents summarized by tag clouds. The tag clouds of titles of documents mentioning Bitcoin on October 3 and 26, 2013 (two peeks in the October occurrence timeline) are presented in Figure 13. It is easy to see that on October 3, 2013 the Silk Road website was shut down and on October 26, 2013, $28 Million in Bitcoin were seized by US prosecutors in the related federal investigation. On the web portal, the titles of the documents are listed below the tag clouds in order of relevance, giving further meaning to the tag clouds.

By using NewsStream, we can relatively easily reconstruct the Bitcoin story. The first Bitcoin ATM in Canada, the US Senate hearing on Bitcoin, Bitcoins being accepted as a tuition payment by a university and Bitcoin reaching the price of $1000 all happened in November 2013, enforcing the Bitcoin price growth and

market capitalization. The most reported news about Bitcoin was the China central bank banning financial companies from using the digital currency in the beginning of December 2013. The European Banking Authority issued a warn on Bitcoin in mid December 2013, followed by the Indian Central Bank at the end of December 2013.

The Bitcoin example is ideal for studying the relation between the financial entity on the market and in the news, since it is, up to now, unregulated. By using the NewsStream portal, we can see that Bicoin first gained a decent market capitalization, followed by mass media attention and, in turn, even bigger demand on exchanges that allow trading Bitcoins for US Dollar.

We have demonstrated the functionality of the NewsStream portal in three diverse use-cases. By starting from an entity (or entities) of interest, NewsStream first answers the questions *when* something was reported about the entity, how big were the stories and what was the associated sentiment. The drill-down to individual events is supported by tag clouds and listings of underlying documents ordered by relevance or sentiment. The NewsStream portal thus gives a novel perspective on the reporting in global news media.

## 3.4 Entity Co-occurence Network Visualization

Within the NewsStream portal we have also implemented a visualization of entity co-occurrences over time. At first we limit ourselves to countries as entities of interest, and represent significant co-occurrences by a network of links between countries. We employ the significance algorithm proposed in [21] to show only links between countries, that did not co-occur in the news by change. The network is embedded into the world map, which enables an intuitive presentation of the relations between countries that significantly co-occur in the news. The temporal dimension can be explored in monthly snapshots, and the significance threshold of the visualized links can be adjusted to views only the most significant links. Figure 14 shows the interactive visualization of the country co-occurrence network.

Tracking entity co-occurrences over time can give many insights into various types of relations between entities, ranging from collaborations or partnerships to conflicts or lawsuits. Entity co-occurrence networks extracted from news have a great potential for exploring the semantics among entities, and identifying groups of entities that co-occur in different contexts. The temporal aspect of the network reveals the emergence and decline of topics in the news, as well as the entities' participation in these topics. Furthermore, the network can be also compared to other real-world networks to discover any structural similarities that might explain how the majority of news is formed.

# 4 Related Work

The analysis of large quantities of news is an emerging area in the field of data analysis and visualization [15]. According to Thomas and Cook (2005) [31],

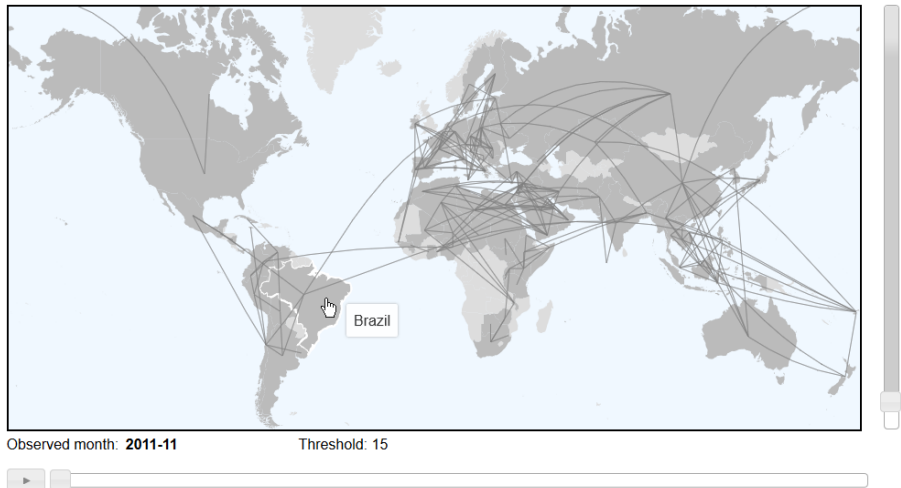

Figure 14: Interactive visualization of the country co-occurrence network in NewsStream.

Visual Analytics is the science of analytical reasoning supported by interactive visual interfaces. In [15], the three challenges of visually representing news streams are identified: First, when dealing with thousands of articles, articles cannot be shown in full detail, and often not even all the titles can be displayed. Second, the timeline approach is inappropriate for some applications, since the old news lose importance and are replaced by the updates or completely different information. Furthermore, keeping all the historic news in the display is infeasible for both, space and performance reasons. Third, in many cases, articles from various sources report on the same topic, producing a lot of redundancy. The NewsStream portal addresses the above challenges by decomposing the analysis into two steps, the information volume and the information content, allowing the user to focus on the desired timespan and content.

Current visual techniques that deal with temporal evolution of such complex datasets, together with research efforts in related domains, such as text mining and topic detection and tracking, represent early attempts to understand, gain insight and make sense of these data. Lack of techniques dealing directly with the problem of visualizing news streams in an 'on-line' fashion is identified in [15]. The authors propose a purely visual technique that permits to see the evolution of news in real-time. A visual analytics system for exploration of news topics in dynamic information streams, which combines interactive visualization and text mining techniques to facilitate the analysis of similar topics that split and merge over time is presented in [17]. CloudLines [14] is a time-series visualization technique of time-based representations of large and dynamic event data sets in limited space. Timeriver [11] visualizes thematic variations over time within a large collection of documents. The 'river' flows from left to right through time, changing width to depict changes in thematic strength of temporally associated documents. In [12], both, the user interaction and technical details for visual analysis of constantly evolving text collections, are described. Our work is most similar to [16], where temporal analysis of occurrence and co-occurrence of entities in news is studied, but there the sentiment is not considered.

Besides the research efforts in the area of news analytics, there are several groups of applications that collect news and blogs information, process them and serve them with added value to the interested users. Here we present the relationship between NewsStream and news aggregation web sites, news media monitoring services and Google Trends. We emphasize similarities and differences to those approaches and present the complementarity and added value of NewsStream.

### 4.1 News Aggregation Websites

News aggregation websites periodically read a set of news sources, find the new bits, group similar stories together, sort the stories according to their importance and display the aggregated information to the user. The aggregation can be entirely automatic (as in Google News, Newslookup, Newsvine, World News Network, and Daily Beast) or semi-automatic, as in the case of JockSpin. A comparison of news aggregator services in terms of their features and usability from the users' perspective is presented in [6].

Similarly to news aggregators, NewsStream's data acquisition pipeline is connected to RSS feeds and regularly pools and processes the data. In contrast to news aggregators which focus on providing a quick overview of the main news of the day, NewsStream provides a historical view on news about selected entities, focusing on their importance and sentiment.

### 4.2 Media Monitoring Services

Europe Media Monitor (EMM, http://emm.newsbrief.eu/) [4, 3] is a news aggregation and analysis system to support EU institutions. It gathers reports from news portals world-wide in 60 languages, classifies the articles, analyses the news by extracting information from them, aggregates the information, issues alerts and produces intuitive visual presentations. It monitors over 10000 RSS feeds and HTML pages from 3750 key news portals world-wide plus 20 commercial news feeds and, for some applications, also specialist sites. It retrieves over 150.000 reports per day in 60 languages. The EMM news gathering engine feeds its articles into automatic public news analysis systems [27]: NewsBrief [28] and MediSys focus on the breaking news and short-term trend detection, early alerting and up-to-date category-specific news display, while NewsExplorer focuses on daily overviews, long-term trends, in-depth analysis and extraction of information about people and organizations.

While the focus of European Media Monitor is to detect events in near real-time, our focus is a comprehensive overview of news about entities over time.

Similarly to NewsExplorer which detects names of persons, organizations and locations, we detect entities from our financial ontology. While NewsExplorer focuses on the description of entities by gathering and summarizing the extracted meta-information on individual news pages, NewsStream focuses on news about entities by showing volumes and aggregated sentiment of news and tag-clouds mentioning those entities over time.

In addition to the European Media Monitor, there are several commercial media monitoring services (also called press clipping services), which provide clients with copies of media content, of specific interest to them. These services tend to specialize their coverage by subject, industry, size, geography, publication, journalist, or editor. PickANews (http://www.pickanews.com/), for example, is a European multimedia search engine indexing over 50,000 media sources (print, web, radio and TV). Results can be analysed by country and type of media. The engine allows users to search for article extracts by keyword, company, brand, product or person. In PickANews, the search and viewing of extracts is free, while subscriptions are required for the alert service by keyword, full content access and visualizing key indicators. Other commercial news monitoring services include MSBA News Clipping Service (http://www.mnmsba.org/NewsClippingService) and WebClipping (http://webclipping.com/).

## 4.3 Google Trends

Google Trends [8] is a public web facility of Google Inc. for displaying search query trends. It shows how often a particular term has been entered into Google Search relative to the total search-volume across various regions of the world, and in various languages (starting from 2004). The results can be further refined by region and time period. Google Trends has weekly data granularity.

The Google Trends data were used in studies in a variety of fields, for example in health to predict the spread of influenza [7] and in finance and economy to relate search query volumes to financial time-series. Search volume data and financial market fluctuations have been investigated by [24] showing that weekly transaction volumes of S&P 500 companies are correlated with weekly search volume of the corresponding company names. Patterns in Google query volumes for search terms related to finance may be interpreted as "early warning signs" of stock market moves [22]. Another study reports a correlation between a country's GDP and the degree to which Internet users worldwide seek more information about years in the future than years in the past [23].

Similar to Google Trends, NewsStream also shows a term occurrence over time. While Google Trends shows the occurrence of terms from search queries, NewsStream shows the occurrence of terms in news articles and financial blog posts. The search query volumes are therefore a complement to media monitoring, as they measure the activity of regular users on the web.

## 5 Conclusions and Further Work

We have presented NewsStream, a visual news analytics platform which focuses on the question when something happened, and what were the amplitude (volume) and sentiment about the event, prior to what happened. The portal allows to focus on specific entities and topics, search through their history, estimate the overall sentiment attributed to those entities through time, evaluate the relative importance of events, summarize collections of news with tag clouds, and observe the evolution of the context in which they appear. It is, therefore, a complementary service to news aggregators and news monitoring services. We have demonstrated the usefulness of NewsStream in four different use cases: (i) study of co-occurrences of rating agencies and the PIIGS countries, (ii) reflection of financial news in CDS prices, (iii) the emergence of the Bitcoin digital currency, and (iv) visualizing how the world is connected through news.

The paper describes the architecture and details about our data acquisition and processing pipeline of (financial) news and blog posts, which acquires, cleans and processes about 40,000 documents (mainstream news and financial blogs) daily in near real time. The underlying database supports efficient querying over specific, finance-oriented aspects of the news.

NewsStream is being further developed in various EU-funded projects. We are working on an inverted index of the documents that will allow querying for arbitrary words in the documents as well as the ontology instances and higher-level terms. We also plan to include topic detection to ease the interpretation of the events on time-lines. Another direction for further work is the generalization to other languages. Large parts of our data acquisition and processing pipeline are language independent (e.g., the boilerplate remover and the duplicate detector), many language dependent components are already developed for many languages (e.g., the lemmatizer and the stop word detector). We have started collecting the news for the Slovenian language as well. Interesting research questions emerge from the availability of such data, for example, the empiric comparison between the visibility and delay of news in its country of origin and in the global news.

To summarize, by making our visual news analytics platform available on the web, we provide an intuitive and easy-to-use use tool for massive news analytics. By putting the questions when something happened, how big was the story, and what was the related sentiment, ahead of what the event was, the NewsStream portal gives a novel view on the reporting in global news media.

## Acknowledgements

This work was supported in part by the European Commission under the FP7 project FOC (Forecasting financial crises, grant no. 255987), and by the Slovenian Research Agency programme Knowledge Technologies (grant no. P2-103). NewsStream is being used, adapted, and further developed in the following EU-funded projects: SIMPOL (Financial Systems Simulation and Policy Modelling, grant no. 610704), MULTIPLEX (Foundational Research on MULTIlevel com-
PLEX networks and systems, grant no. 317532), and DOLFINS (Distributed
Global Financial Systems for Society, grant no. 640772). We also thank Sašo
Rutar and Matjaž Juršič for their help in solving various technical and concep-
tual issues.